\documentstyle[12pt]{article}
\newcommand{\nc}{\newcommand}

\nc{\dbar}{\bar{\partial}}

\catcode`\@=11

\@addtoreset{equation}{section}
\def\theequation{\thesection\arabic{equation}}

\def\@normalsize{\@setsize\normalsize{15pt}\xiipt\@xiipt
\abovedisplayskip 14pt plus3pt minus3pt%
\belowdisplayskip \abovedisplayskip
\abovedisplayshortskip  \z@ plus3pt%
\belowdisplayshortskip  7pt plus3.5pt minus0pt}
\def\small{\@setsize\small{13.6pt}\xipt\@xipt
\abovedisplayskip 13pt plus3pt minus3pt%
\belowdisplayskip \abovedisplayskip
\abovedisplayshortskip  \z@ plus3pt%
\belowdisplayshortskip  7pt plus3.5pt minus0pt
\def\@listi{\parsep 4.5pt plus 2pt minus 1pt
            \itemsep \parsep
            \topsep 9pt plus 3pt minus 3pt}}

\def\underline#1{\relax\ifmmode\@@underline#1\else
        $\@@underline{\hbox{#1}}$\relax\fi}
\@twosidetrue
\relax

\catcode`@=12

\catcode`\@=11

\def\section{\@startsection{section}{1}{\z@}{3.5ex plus 1ex minus
   .2ex}{2.3ex plus .2ex}{\large\bf}}

\def\ps@headings{\def\@oddfoot{}\def\@evenfoot{}
\def\@oddhead{\hbox{}\hfill
        \makebox[.5\textwidth]{\raggedright\ignorespaces --\thepage{}--
        \hfill }}
\def\@evenhead{\@oddhead}
\def\subsectionmark##1{\markboth{##1}{}}
}

\ps@headings

\catcode`\@=12

\relax

\def\figcap{\section*{Figure Captions\markboth
        {FIGURECAPTIONS}{FIGURECAPTIONS}}\list
        {Fig. \arabic{enumi}:\hfill}{\settowidth\labelwidth{Fig. 999:}
        \leftmargin\labelwidth
        \advance\leftmargin\labelsep\usecounter{enumi}}}
 \relax
\def\tablecap{\section*{Table Captions\markboth
        {TABLECAPTIONS}{TABLECAPTIONS}}\list
        {Table \arabic{enumi}:\hfill}{\settowidth\labelwidth{Table 999:}
        \leftmargin\labelwidth
        \advance\leftmargin\labelsep\usecounter{enumi}}}
 \relax
\def\reflist{\section*{References\markboth
        {REFLIST}{REFLIST}}\list
        {[\arabic{enumi}]\hfill}{\settowidth\labelwidth{[999]}
        \leftmargin\labelwidth
        \advance\leftmargin\labelsep\usecounter{enumi}}}
 \relax

\catcode`\@=11

\def\ps@headings{\def\@oddfoot{}\def\@evenfoot{}
\def\@oddhead{\hbox{}\hfill
        \makebox[.5\textwidth]{\raggedright\ignorespaces --\thepage{}--
        \hfill }}
\def\@evenhead{\@oddhead}
\def\subsectionmark##1{\markboth{##1}{}}
}

\ps@headings

\relax

\def\firstpage#1#2#3#4#5#6{
\begin{document}

\begin{titlepage}
\nopagebreak
\title{\begin{flushright}
      \vspace*{-1.3in}
        {\normalsize TUM--HEP-262/96}\\[-5mm]
        {\normalsize hep-th/9611110}\\[-5mm]
{\normalsize November 1996}\\[.5cm]
\end{flushright}
\vspace{20mm}
{\large \bf #3}}
\vspace{1cm}
\author{\large #4 \\ #5}
\maketitle
\vskip -1mm
\nopagebreak
\begin{abstract}
{\noindent #6}
\end{abstract}
\vspace{3cm}
\begin{flushleft}
\rule{16.1cm}{0.2mm}\\[-3mm]
$^{\star}${\small Work supported by the European Commission
TMR programmes ERBFMRX-CT96-0045
and \\[-2mm] ERBFMRX-CT96-0090.}\\[-2mm]
$^{1}${\small Supported by  the  Alexander von Humboldt--Stiftung}\\[-2mm]
$^{2}${\small E--mail: kehagias@physik.tu-muenchen.de}
\end{flushleft}
\thispagestyle{empty}
\end{titlepage}}
\newcommand{\dal}{\raisebox{0.085cm}
{\fbox{\rule{0cm}{0.07cm}\,}}}
\newcommand{\bb}{\begin{eqnarray}}
\newcommand{\ee}{\end{eqnarray}}
\newcommand{\p}{\partial}
\newcommand{\bp}{{\bar \p}}
\newcommand{\bR}{{\bf R}}
\newcommand{\bC}{{\bf C}}
\newcommand{\bZ}{{\bf Z}}
\newcommand{\bS}{{\bar S}}
\newcommand{\bT}{{\bar T}}
\newcommand{\bU}{{\bar U}}
\newcommand{\bA}{{\bar A}}
\newcommand{\bh}{{\bar h}}
\newcommand{\bu}{{\bf{u}}}
\newcommand{\bv}{{\bf{v}}}
\newcommand{\D}{{\cal D}}
\newcommand{\s}{\sigma}
\newcommand{\Sg}{\Sigma}
\newcommand{\ket}[1]{|#1 \rangle}
\newcommand{\bra}[1]{\langle #1|}
\newcommand{\non}{\nonumber}
\newcommand{\ph}{\varphi}
\newcommand{\la}{\lambda}
\newcommand{\ga}{\gamma}
\newcommand{\ka}{\kappa}
\newcommand{\m}{\mu}
\newcommand{\n}{\nu}
\newcommand{\th}{\vartheta}
\newcommand{\Lie}[1]{{\cal L}_{#1}}
\newcommand{\eps}{\epsilon}
\newcommand{\bz}{\bar{z}}
\newcommand{\bX}{\bar{X}}
\newcommand{\om}{\omega}
\newcommand{\Om}{\Omega}
\newcommand{\we}{\wedge}
\newcommand{\La}{\Lambda}
\newcommand{\bOm}{{\bar \Omega}}
\newcommand{\CA}{{\cal A}}
\newcommand{\CF}{{\cal F}}
\newcommand{\CbF}{\bar{\CF}}
\newcommand{\CAM}{\CA^{(M)}}
\newcommand{\CAS}{\CA^{(\Sg)}}
\newcommand{\CFS}{\CF^{(\Sg)}}
\newcommand{\I}{{\cal I}}
\newcommand{\al}{\alpha}
\newcommand{\be}{\beta}
\newcommand{\cm}{Commun.\ Math.\ Phys.~}
\newcommand{\pr}{Phys.\ Rev.\ D~}
\newcommand{\pl}{Phys.\ Lett.\ B~}
\newcommand{\ibar}{\bar{\imath}}
\newcommand{\jbar}{\bar{\jmath}}
\newcommand{\np}{Nucl.\ Phys.\ B~}
\newcommand{\e}{{\rm e}}
\newcommand{\gsi}{\,\raisebox{-0.13cm}{$\stackrel{\textstyle
>}{\textstyle\sim}$}\,}
\newcommand{\lsi}{\,\raisebox{-0.13cm}{$\stackrel{\textstyle
<}{\textstyle\sim}$}\,}
\date{}
\firstpage{95/XX}{3122}
{\Large\sc N=2 Heterotic
Stringy Cosmic Strings$^{\star}$} { Alexandros Kehagias$^{1,2}$}
{\normalsize
Physik Department\\
\normalsize Technische Universit\"at M\"unchen \\
\normalsize D-85748 Garching, Germany}
{ We construct  solitonic string solutions
of  N=2 four--dimensional heterotic models
of rank three, four and five.
These finite energy configurations have constant dilaton while the
moduli fields vary over space--time with jumps at the location of the
string cores consistent
with the T--duality groups $SL(2,\bZ),SL(2,\bZ)\!\times\!SL(2,\bZ)$
and $Sp(4,\bZ)$.
The solutions are  expressed in terms of
modular forms of the T--duality group.  They break half of the
supersymmetries and the vacuum contain a certain
number of solitonic strings in order the singularities to be resolved
in a Ricci flat way.}

\newpage

\section{Introduction}

Theories with extended supersymmetries reveal a rich dynamical structure
\cite{w0}--\cite{w1}. An
important feature of these theories is the existence of BPS states which
break half of the supersymmetries. They are of particular importance in
determine the dynamics of the theory and their semi-classical analysis in some
cases is enough to determine also their strong coupling behaviour.
On the other hand, they are necessary for the consistency of the
theory. For example, BPS states which carry Ramond--Ramond
charge are the D--branes \cite{pol}
of type II theory where the type I string may have its
ends. They also play a central role in establishing the various string/string
dualities,   in M-- and F--theory,
in various string compactifications and so on
\cite{w1},\cite{ht}--\cite{vaf}.
It seems that all the  recent developments indicate
that the understanding of the structure of the BPS states will be a central
issue in unrevealing the secrets of string theory.

In this paper, we will deal with four--dimensional
supersymmetric  N=2 models and we will construct string--like
configurations which satisfy a Bogomol'nyi bound and break half of the
space--time supersymmetries.
N=2 supersymmetry in four dimensions can be obtained by
compactifying type II strings on a Calabi--Yau threefolds with Betti
numbers $h_{11}$ and $h_{12}$. The  number $h_{11}$
in type IIA theory for example, gives the
number of vector multiplets and combined with the
graviphoton, the rank of the gauge group turns out to be
$h_{11}+1$. On the other
hand, the dilaton belongs to a hypermultiplet and the total
number of hypermultiplets is $h_{12}+1$. The tree level prepotential
is exact in the full quantum theory
since the dilaton belongs to a hypermultiplet and
the mirror symmetry made possible its exact computation
\cite{can}--\cite{cand}.
There exist in general logarithmic singularities near
the conifold locus in the moduli space of the CY threefold
\cite{cand}. The
problem of these conifolds singularities resolved by Strominger who
proposed that hypermultiplets corresponding to charged black holes become
massless near the conifold locus \cite{strom}.

The compactification of the heterotic string on $K_3\times T^2$ also
give rise to N=2 supersymmetry in four dimensions with
 $n_V+1$ vector multiplets, including the graviphoton,
and $n_H$ hyper multiplets. The rank of the
gauge group is $n_V+2$ in this case since the dilaton belongs to a vector
multiplet now. Supersymmetry requires the moduli space of the vector
multiplets to be a special K\"ahler manifold ${\cal K}_{n_V}$ while the
moduli space of the hypermultiplets is a quartenionic manifold
${\cal Q}_{n_H}$ \cite{K1},\cite{K2}.
In fact, the vector multiplets parametrize the
coset $\frac{SU(1,1)}{U(1)}\times\frac{O(2,n_V)}{O(2)\times~O(n_V)}$
where the first factor is the moduli space of the S--dilaton. The
classical T--duality group is $O(2,n_V;\bZ)$ which, however,
is  modified by quantum corrections \cite{and}.

Here, we will consider  string-like soliton
solutions to the low-energy four-dimensional effective action
which break half of the supersymetries. We discuss first an  N=4 model with
dilaton $S$ and two additional moduli $T$ and $U$. The major difficulty in
constructing solitonic string  configurations
comes from the fact that, without any extra assumption,
all of them have infinite energy
per unit length which
leads to inconsistencies in solving the field equations
\cite{g}. However, one may construct finite energy
solutions  by employing the
S-- and T--duality groups and restricting the moduli fields to their
fundamental domain. The moduli space has then finite volume and
the energy turns out to be proportional to it.

In the case of  N=2
heterotic strings in four--dimensions we explicitly discuss the
rank three, four and five  models with T--duality groups $SL(2,\bZ)$,
$SL(2,\bZ)_T\times SL(2,\bZ)_U$ and $Sp(4,\bZ)$, respectively. These
models have also been discussed in connection with heterotic/type IIA
string duality \cite{kv}--\cite{lust}.
Here again, the string configurations have infinite energy and
finite energy solutions can only be constructed  if one
allows the moduli fields to have
discontinuous jumps as they
go around the string as long as these jumps have been
done by an element of the T-duality group.  In other words,
solitonic solutions exist only if the fields are restricted to
the fundamental domain of the T--duality group.
One  should  recall at this point the
stringy cosmic string of Greene et. al. where the $SL(2,{\bf Z})$ T-duality
group of a torus compactifications was been employed in order finite energy
solutions to be constructed for a single moduli \cite{g}. It should be
noted that we fix the value of the $S$--dilaton since the S--duality group is
lacking in the N=2 case.

String--like configurations which break half of the supersymmetries
have previously been exploited as well. In \cite{dab}, for example, a
multi--string solution of the three--form ten--dimensional
supergravity coupled to a string $\sigma$--model source has been
constructed. This solutions  was shown to satisfy a
Bogomol'nyi bound and to break half of the space--time
supersymmetries. Strictly speaking,
it is not a genuine soliton since requires the presence of an
``electrically" charged source
due to a singularity at the location of the string. However, it can be
interpreted as a soliton of the fivebrane theory. Other genuine
solitonic string solutions have also been constructed \cite{sol} and
a review of the subject can be found in \cite{duff}.

In the next chapter, we present the general setting of our
constructions by discussing the solitonic string solutions in
$\sigma$--models coupled to gravity. We also recall the Greene et al.
solutions and we find stringy cosmic string solutions in the  N=4 heterotic
theory with three moduli $S,T,U$ in four dimensions. In chapter 3
we construct
solitonic string solutions of the rank three, four and five $S-T$,
$S-T-U$ and $S-T-U-V$ models. In chapter 4 we discuss some issues of
our solutions and finally, in an appendix we present some properties
of the Siegel modular group $Sp(4,\bZ)$.

\section{Solitonic string solutions}
\subsection{General setting}

The number of supersymmetries after compactification down to
four dimensions is determined by the number of covariantly
constant spinors in the internal six--dimensional space.
For example, a Calabi--Yau
compactification give rise to N=1 (2) supersymmetry
while a $K_3\times T^2$ or a $T^6$ compactification give N=2 (4) and N=4 (8)
supersymmetry in four dimensions for heterotic (type II)
strings.
The form of the  moduli
space is then restricted by supersymmetry and for the N=2 case
 turns out to be
${\cal M}={\cal K}\times {\cal Q}$ where
${\cal K}$ is a K\"ahler manifold for the moduli space of
the vector multiplets and
 ${\cal Q}$ is quaternionic for the hyper multiplets.

Let us consider  the universal part of the  effective
action of the N=2 four--dimensional heterotic string which describes
the dynamics of the graviton and the scalar components of the vector
multiplets. The moduli space of the latter is
a special  K\"ahler manifold ${\cal
K}$ with local coordinates $(w^i,\bar{w}^i;
i=1,\cdots,\dim_\bC{\cal K})$. The metric on ${\cal K}$ is
$h_{i\bar{j}}=\partial_i\partial_{\bar{j}} K(w,\bar{w})$
where $K(w,\bar{w})$ is the K\"ahler potential. The
bosonic part of the
one--loop corrected effective action up to first order in
$\alpha^\prime$--expansion is
\bb
I&=&\int d^4 x\sqrt{-g} \left( \frac{1}{2}R-
h_{i\bar{j}}\partial_\m
w^i\partial^\m \bar{w}^j+\frac{1}{8}S_2
R_{GB}^2+\frac{1}{8}S_1RR^\ast\right. \nonumber \\
&&\left.+\Delta(w^i,\bar{w}^i)
R_{GB}^2+\Theta(w^i,\bar{w}^i)RR^\ast\right) \, , \label{action}
\ee
where $\Delta(w^i,\bar{w}^i)$, $\Theta(w^i,\bar{w}^i)$ are
 the moduli--dependent one--loop
corrections \cite{andd},
$S_1$, $S_2$ are the real and imaginary parts of
the S--dilaton and $R_{GB}^2$, $RR^\ast$ are the CP--even Gauss--Bonnet
combination and the CP--odd term defined by
\bb
R_{GB}^2&=&R_{\kappa\lambda\m\n}R^{\kappa\lambda\m\n}-4R_{\m\n}R^{\m\n}
+R^2\, ,\nonumber \\
RR^\ast&=&\epsilon^{\kappa\lambda\m\n} {R_{\kappa\lambda}}^{\alpha\beta}
R_{\m\n\alpha\beta} \, .
\ee
We are looking for solitonic  string-like solutions of the form
\bb
ds^2=-dt^2+dx_3^2+e^{\rho(z,\bar{z})} dzd\bar{z}\, ,\label{metr}
\ee
where the complex coordinates $(z,\bar{z})$ parametrize the plane
transverse to the string which is  extended in the $x_3$-directions
and the complex moduli are $w^i=w^i(z,\bar{z})$.
With this form of the metric, both the Gauss--Bonnet and the CP--odd terms
vanish and it is consistent with the equations of motions to ignore them.
Thus, the effective actions takes the form of a $\sigma$--model
coupled to gravity
\bb
I=\int d^4 x\sqrt{-g} \left( \frac{1}{2}R-
h_{i\bar{j}}\partial_\m
w^i\partial^\m \bar{w}^j\right) \, , \label{action0}
\ee
and the equations of motions are then
\bb
R_{\m\n}&=&
h_{i\bar{j}}\partial_\n
w^i\partial_\m \bar{w}^j +
h_{i\bar{j}}\partial_\m w^i \bar{w}^j \, , \label{einstein}\\
0&=&\frac{1}{\sqrt{-g}}\partial_\m\left(g^{\m\n}\sqrt{-g}\partial_\n
h_{i\bar{j}}w^i\right)-h_{i\bar{k},\bar{j}}\partial_\m
w^i\partial^\m \bar{w}^j \, , \label{w}
\ee
where $h_{i\bar{k},\bar{j}}=\partial h_{i\bar{k}}/\partial \bar{w}^j$.

The four dimensional  space-time
is of the form ${\bf R}^2\!\times\!\Sigma$ where
$\Sigma$ is a two-dimensional surface with  Euler number
\bb
\chi(\Sigma)=-\frac{i}{2\pi}\int d^2z\partial\bar{\partial}\rho
\, .\label{Euler}
\ee
The energy  per unit length of such configurations in complex notation is
\bb
E=\frac{i}{2}
\int d^2zh_{i\bar{j}}\left(\partial w^i\bar{\partial}
\bar{w}^j+\partial w^i\partial \bar{w}^j\right) \, , \label{energy}
\ee
and it is easy to verify that it satisfies the BPS bound
\bb
E \geq
\left|\frac{i}{2}\int d^2zh_{i\bar{j}}\left(\partial w^i\bar{\partial}
\bar{w}^j-\partial w^i\partial \bar{w}^j\right)\right| \, . \label{bound}
\ee
The BPS saturated states are then holomorphic (anti-holomorphic)
functions  $w^i=w^i(z)$ $\left(w^i=w^i(\bar{z})\right)$ with  energy per unit
length
\bb
E=\frac{i}{2}\int d^2z h_{i\bar{j}}\partial w^i\bar{\partial}\bar{w}^j
\, . \label{BPS}
\ee
By recalling that
$h_{i\bar{j}}=\partial_i\partial_{\bar{j}} K$ we may express  the energy
in terms of the
K\"ahler potential $K$ as
\bb
E=\frac{i}{2}\int_{w(\Sigma)}\partial\bar{\partial} K(w,\bar{w}) \, ,
\label{energykahh}
\ee
where $w(\Sigma)$ is the image of $\Sigma$ in ${\cal K}$ and here
$(\partial,\bar{\partial})$ are Dolbeault operators.
Although E looks to be a total derivative and thus it should be zero in
the compactified z-plane,  it is not, since
the K\"ahler potential $K$ is not a globally defined quantity.
We will verify this later when we will explicitly calculate the
integral in eq.(\ref{energykahh}).

The equations for $w^i,\bar{w}^i$ are automatically satisfied if the
BPS condition is fulfilled. One may verify that holomorphic
or antiholomorphic $w^i$ indeed solves eq.(\ref{w}).
Thus, only the Einstein equations eq.(\ref{einstein})
remain to be solved. They turn out to be the single equation
\bb
\partial\bar{\partial}\rho=-h_{i\bar{j}}\partial
w^i\bar{\partial}\bar{w}^j\, , \label{KK}
\ee
for the conformal factor $\rho(z,\bz)$. In terms of
the K\"ahler potential $K$, eq.(\ref{KK}) is written as
\bb
\partial\bar{\partial}\rho=-\partial\bar{\partial}K
\, . \label{K}
\ee
The solutions to eq.(\ref{K}) is expressed in terms of an
arbitrary  holomorphic faction $F(z)$ as
\bb
\rho(z,\bar{z})=-K(w,\bar{w})+F(z)+F(\bar{z})
\, . \label{solution}
\ee
Thus,
the metric for the static cylindrically symmetric space-time turns out to be
\bb
ds^2=-dt^2+dx_3^2+e^{-K(w,\bar{w})}|h(w)|^2dzd\bar{z} \, ,
\label{metricsol}
\ee
where, by taking into account the holomorphicity of the field $w$ we
have written $h(w)=\exp{F(z)}$. The holomorphic function $F(z)$ or $h(w)$
can be specified by demanding non--degenerate metric.
Moreover, by comparing
eqs.(\ref{Euler},\ref{BPS},\ref{KK}), the energy per unit length
is expressed in
terms of the Euler number of $\Sigma$ as
\bb
E=2\pi\chi(\Sigma) \, . \label{Eulerene}
\ee

As an explicit example,
let us consider the case of an $SU(2)/U(1)$ $\sigma$-model \cite{Gom}
with K\"ahler
potential, in projective coordinates ($w,\bar{w}$), $K=2n\log (1+w\bar{w})$
(the factor  $2n$ in front
is necessary for the scalar manifold to be  a Hodge
manifold \cite{WB}).
Then the metric (\ref{metricsol}) turns out to be
\bb
ds^2=-dt^2+dx_3^2+\frac{|h(w)|^2}{(1+w\bar{w})^{2n}}dzd\bar{z} \, .
\label{metricsol1}
\ee
Finite energy solutions are provided by the instanton configurations
\bb
w(z)=\sum_{i=1}^{N} \frac{z-a_i}{z-b_i}\, , \label{ww}
\ee
and thus,  the metric vanishes as $|h(w)|^2\prod_{i=1}^N|z-b_i|^{4n}$.
The condition
of a nowhere vanishing metric leads to the choice
 $h(w)=1/\prod_i(z-b_i)^{2n}$ so that we get
\bb
ds^2=-dt^2+dx_3^2+\frac{1}{(1+w\bar{w})^2\prod_{i=1}^N|z-b_i|^{4n}}
dzd\bar{z} \, .
\label{metricsol2}
\ee
The energy per unit length of these configurations is $E=2\pi nN$.
At infinity ($|z|\rightarrow \infty$), the metric of the transverse
space goes like  $e^\rho\sim 1/|z|^{4nN}$ and thus there exists a
deficit angle $\delta=2\pi nN$. We recall that a deficit angle of $2\pi$
corresponds to an asymptotically
cylindrical space, a deficit angle greater than
$2\pi$ to a conical space with infinity in finite distance while a
deficit angle of $4\pi$ corresponds to $\bf{ CP}^1$. Put differently, a
deficit angle  $2\pi$ ($4\pi$) corresponds to a surface with
Euler number $\chi=1$ ($\chi=2$)). Thus, if $n=1$, two--string
configurations  compactify the transverse space on ${\bf CP}^1$
while if $n=2$, one string is enough to close it up.

\subsection{Stringy cosmic strings}

We will describe here the stringy cosmic string solution of Greene et
al. \cite{g} which is the  prototype  of solutions we are going to construct.
Let us consider the $SL(2,{\bf R})/U(1)$
$\sigma$-model coupled to gravity in 4-dimensions
which is obtained after dimensional reduction of the
6-dimensional Einstein gravity on a
torus $T^2$. If one fixes the volume of $T^2$ to some constant
value, the only massless moduli
is then a complex scalar field which is
the complex structure modulus $\tau$ of the torus.
The target space ${\cal{M}}=SL(2,{\bf R})/U(1)$ is the upper half plane
${\cal H}_1$ which  is  K\"ahler  with
K\"ahler potential $K=-\log \tau_2$ ($\tau_2=Im\tau>0$).
The bosonic part of the low energy
effective action is
\bb
I=\int d^4x \sqrt{-g}\left(\frac{1}{2}R+
\frac{\partial_\m \tau \partial^\m \bar{\tau}}{(\tau-\bar{\tau})^2}\right)
\, , \label{actionS}
\ee
and it is invariant under the global $SL(2,\bR)$ transformations
\bb
\tau\rightarrow \frac{a\tau+b}{c\tau+d}\, , ~~~~~~~~~
\left(\begin{array}{cc}
a&b\\c&d\end{array}\right) \in SL(2,\bR) \, .
\ee
By identifying the complex scalar $\tau$ with the complex structure
moduli of the internal torus,  the theory is invariant under
the modular group $PSL(2,\bZ)=SL(2,\bZ)/{\bf Z_2}$ since
$PSL(2,\bZ)$ transformations of $\tau$ give back the same
torus. The generators of the modular group  are   the
transformations $\tau\rightarrow \tau+1$ and $\tau\rightarrow
-1/\tau$.
The equations of motion are now
\bb
R_{\m\n}&=&-
\frac{\partial_\m \tau\partial_\n \bar{\tau}}{(\tau-\bar{\tau})^2}
-\frac{\partial_\n \tau\partial_\m \bar{\tau}}{(\tau-
\bar{\tau})^2}\, , \label{Ein}
\\
0&=& \frac{1}{\sqrt{-g}}\partial_\m\left(g^{\m\n}\sqrt{-g}
\partial_\n \tau\right)+
2\frac{\partial_\m \tau\partial^\m \tau}{\tau-\bar{\tau}}
\, , \label{eqS}
\ee
and the stringy cosmic string will be described by a metric of the form
eq.(\ref{metr}). The equation for $\tau=\tau(z,\bar{z})$
turns out then to be
\bb
\p\bp\tau+2\frac{\p\tau\bp\tau}{\tau-\bar{\tau}}=0,
\ee
and it is solved for holomorphic or antiholomorphic field. We will assume that
$\tau=\tau(z)$ in the following.  The energy per unit length
according to eq.(\ref{energykahh}), is then
\bb
E=-\frac{i}{2}\int  \partial\bar{\partial}\ln \tau_2 \, , \label{enS}
\ee
and it  diverges.
In order to find finite energy solutions
one has to restrict $\tau$ to the fundamental domain
of $PSL(2,\bZ)$~\cite{g}. Then, $\tau$ has discontinuous jumps
done by the PSL(2,Z) transformations  $\tau\rightarrow \tau+1$
as we go around the string.  These jumps  and the holomorphicity
require that near the location of the string
\bb
\tau\simeq \frac{1}{2\pi i}\ln z \, .  \label{tz}
\ee
The energy in this case is indeed finite and it turns out to be proportional to
the volume of the fundamental domain ${\cal F}_1$,
\bb
E=\frac{\pi}{6}n \, , \label{ener}
\ee
where $n$ is the number of  times
the z-plane covers ${\cal F}_1$.

Since the fundamental domain  of $SL(2,\bZ)$ is mapped to
the complex sphere in the j--plane through the modular j-function, we
may express the solution for $\tau$ as the pull-back of  $j(\tau)$.
Thus we may write
\bb
j(\tau)=\frac{P(z)}{Q(z)} \, ,
\ee
where $P(z),Q(z)$ are polynomials of degree $p$ and $q$, respectively.
If $p\leq q$, $j$ approaches a constant value as $|z|\rightarrow \infty$
and $n=q$ in this case. There exist q points at which $Q(z)$ has
zeroes and these points may be considered as the locations of the string
cores. On the other hand, if $p>q$, the solution diverges at
$|z|\rightarrow \infty$ and $n=p$ now.

Turning  to the Einstein equations eqs.(\ref{Ein}),
only the (00) equation  is not
automatically satisfied and it is  written as
\bb
\partial\bar{\partial}\rho=
\frac{\partial \tau\bar{\partial}\bar{\tau}}
{(\tau-\bar{\tau})^2}\, . \label{rho}
\ee
By recalling the general discussion of the previous section or
by an explicit calculation,
one may easily verify that eq.(\ref{rho}) is solved by
\bb
\rho=\tau_2 |h(\tau)|^2 \, ,
\ee
where $h(\tau)$ is an arbitrary holomorphic function.
The latter can be specified by demanding
non-degenerate metric as well as  modular invariance. These
two conditions give
the supersymmetric solution
\bb
e^{\rho}=\tau_2\eta(\tau)^2\bar{\eta}(\bar{\tau})^2
\left|\prod_{i=1}^{n}(z-z_i)^{-1/12}\right|^2 \, , \label{Eins}
\ee
where $\eta(\tau)=q^{1/24}\prod_{r>0}(1-q^r)$ is the
Dedekind's $\eta$-function ($q=e^{2\pi i\tau}$).
The asymptotic form of the space-time metric is then
\bb
ds^2\sim -dt^2+dx^3+(z\bar{z})^{-n/12}dzd\bar{z} \, ,
\ee
and one recognizes  a deficit angle $\delta=\pi n/6$.
With $n=12$ strings the deficit angle becomes $\delta=2\pi$ and the transverse
space is asymptotically a cylinder while $n=24$ strings produce a deficit angle
$\delta=4\pi$ and the transverse space is a compact ${\bf CP}^1$.

Before closing this section, let us also note that this solution is
also the prototype of the seven-brane solution \cite{gg}.
In the latter case, the modulus
$\tau$ corresponds to the ten-dimensional axion-dilaton field of
type IIB theory while the $PSL(2,\bZ)$ symmetry to strong--weak
coupling duality. The  seven-branes break half of the
space-time supersymmetries and 24 of them compactify the
transverse space on ${\bf CP}^1$. This configuration may then be viewed as a
consistent type IIB vacuum. If in addition, one identifies the
axion-dilaton field  with the complex structure modulus $\tau$ of a torus
compactification of a twelve--dimensional theory \cite{hull+},
the 24 seven-brane
configuration of type IIB theory corresponds to a $K3$
compactification of the twelve dimensional F--theory \cite{vaf}.

\subsection{Four-dimensional N=4 stringy cosmic strings}

The effective action for N=4 supergravity can be obtained by dimensional
reduction on a six torus of the ten-dimensional N=1 supergravity
coupled to N=1 super Yang--Mills theory \cite{sup}.
If one restrict himself in
the  $U(1)^{16}$ part of the  gauge
group in ten dimensions, which is also that part that will give rise
to massless moduli, the four-dimensional action is \cite{w1}
\bb
I_{H}&=&\int d^4xe^{-2\phi}\left(R+4\p_\m \phi\p^\m \phi-
\frac{1}{12}H_{\m\n\ka}H^{\m\n\ka}\right.\nonumber \\
&&\left. -\frac{1}{4}
F_{\m\n}^I(LML)_{IJ}F^{J\m\n}+\frac{1}{8}Tr(\p_\m ML\p^\m ML)\right)\, ,
\ee
where $F_{\m\n}^I=\p_\m A_\m^I-\p_\n A_\m^I,
(I=1,\cdots,28),\,\,
H_{\m\n\ka}=\p_\m B_{\n\ka}+2A_\m^I L_{IJ}
F_{\n\ka}^J + cyclic\,\, perm. \, ,$
and  M is a $(28\times28)$--matrix which satisfies
\bb
MLM^T=L\, , \hspace{.2cm} M^T=M\, , \hspace{.2cm}
L=\left(\begin{array}{ccc}
0&I_6&0\\
I_6&0&0\\
0&0&-I_{16}
\end{array}\right) \, .
\ee
The entries of  $M$ are expressed in terms of  the  scalars of the
theory which
parametrize the coset $O(6,22)/O(6)\times O(22)$ \cite{nar}.
The effective action is invariant under the $O(6,22)$ transformations
\bb
M\rightarrow \Omega M\Omega^T\, , ~~~~~~~~A_\m\rightarrow \Omega A_\m
\, ,
\ee
which leave all other fields invariant and where $\Omega$ is an
$O(6,20)$  matrix satisfying $\Omega^TL\Omega=L$.

We will
consider here only an $O(2,2)/O(2)\times O(2)$
subspace of the full moduli space
which can be obtained from six dimensions as follows.
By toroidal compactification of the
ten--dimensional  N=1 supergravity we get N=2 supegravity
in six dimensions with  moduli
space  $O(4,20)/O(4)\times O(20)$. Further compactification on a two torus will
give N=4 in four-dimensions with moduli space
 \bb
\frac{O(2,2)}{O(2)\times O(2)} \times \frac{O(4,20)}{O(4)\times O(20)}\, ,
\label{ooo}
\ee
if there are no components of the six--dimensional gauge fields
along the two torus. We will consider only the
first factor of (\ref{ooo}) and we will fix all other moduli
to some constant value. Combined with the dilaton S--field,
we will deal in the following with
\bb
{\cal M}=\frac{SL(2,\bR)}{U(1)}\times\frac{O(2,2)}{O(2)\times O(2)}\,
,  \label{soo}
\ee
and  the only non-vanishing scalars  will be
the dilaton S and the  K\"ahler  and complex
structure moduli of the  torus T and U, respectively.
They are defined as
\bb
S&=&\alpha+ie^{-2\phi}\, , \nonumber \\
T&=&B_{45}+i\sqrt{detG_{mn}} \, , \nonumber \\
U&=&\frac{G_{45}}{G_{55}}+i\frac{\sqrt{detG_{mn}}}{G_{55}}  \, ,
\label{def}
\ee
where $G_{mn} (m,n=4,5)$ and $B_{45}$ are  the metric
and the component of the antisymmetric tensor on the torus and
$\alpha, \phi$ are
the axion and the dilaton, respectively.
 The bosonic part of the action is then
\bb
I=\int d^4x \sqrt{-g}\left(\frac{1}{2}R-
\frac{\partial_\m S \partial^\m \bar{S}}{(S-\bar{S})^2}-
\frac{\partial_\m T \partial^\m \bar{T}}{(T-\bar{T})^2} -
\frac{\partial_\m U \partial^\m \bar{U}}{(U-\bar{U})^2}\right)
\, . \label{STU}
\ee
The T-duality group is $SL(2,\bZ)_T\times
SL(2,\bZ)_U$  and in addition, the theory is believed to
be also  invariant under the S-duality group $SL(2,\bZ)_S$.
These discrete groups act on the fields as
\bb
S\rightarrow \frac{aS+b}{cS+d}\, ,\hspace{.2cm}
\left(\begin{array}{cc}
a&b\\c&d\end{array}\right)\in SL(2,\bR)_S\, ,
\ee
and similarly for $T,U$.
There is also
another discrete symmetry, the  string/string/string triality
which  interchanges $S\leftrightarrow T\leftrightarrow U$ \cite{dd}.
Although part of the triality is realized on-shell, here is manifest at
the level of the action since we have turned off all gauge fields.

Solitonic string solutions  of the form  eq.(\ref{metr})
can be constructed by recalling that the prepotential is
$F=STU$ and the K\"ahler potential $K=-\log(S_2T_2U_2)$ ($S_2,T_2,U_2$
are the imaginary parts of the $S,T,U$ moduli). Then,
from eq.(\ref{metricsol}) it follows that the metric is given by
\bb
ds^2=-dt^2+dx_3^2+S_2T_2U_2|f(z)|^2dzd\bz \, . \label{met}
\ee
The moduli fields $S,T,U$ are holomorphic and, as before,
finiteness of the  energy is achieved by restricting them on the
fundamental domains of $SL(2,\bZ)_S,\, SL(2,\bZ)_T$ and $SL(2,\bZ)_U$,
respectively. In this case, the solution may be expressed as the
pull--backs of $j(S),\, j(T)$ and $j(U)$ and $S,T,U$ will be given by
\bb
S&\sim& \frac{1}{2\pi i}\log (z-z_k)\, ,~~~~~~~\\
T&\sim& \frac{1}{2\pi i}\log (z-z_i)\, ,~~~~~~~\\
U&\sim& \frac{1}{2\pi i}\log (z-z_j)\, ,
\ee
near the core of the strings. The conditions of modular
invariance and non--degeneracy of the metric give now the solution
\bb
ds^2=-dt^2+dx_3^2+S_2T_2U_2|\eta(S)\eta(T)\eta(U)|^4
\left|\prod_{i=1}^{n_S}
\prod_{j=1}^{n_T}
\prod_{k=1}^{n_U}
(z-z_i)(z-z_j)(z-z_k)\right|^{-1/6} \!\!\!\!\!\!\!dzd\bz\, , \label{solSTU}
\ee
where $n_S,n_T,n_U$ are the number of strings carrying $S,T$ and $U$
charge, respectively. The total energy is  $E=(n_S+n_T+n_U)\pi/6$.
Finally, string/string/string triality \cite{dd}
requires $n_S=n_T=n_U=n$ while
from the asymptotic behaviour of the metric it turns out that $n=8$ in
order the transverse space to be ${\bf CP}^1$.

The form of the metric (\ref{solSTU}) indicates that there exist eight
S--strings, strings which
carry S--charge, eight T--strings  and eight U--strings.
This configuration compactifies the transverse space on  ${\bf CP}^1$.
On the other hand, string/string/string triality allows also for
STU--strings, that is strings which
carry both S,T and U charges. The transverse space
metric for these strings is
\bb
ds_\bot=S_2T_2U_2|\eta(S)\eta(T)\eta(U)|^4
\left|\prod_{i=1}^{8}
(z-z_i)^{-1/4}\right|^{2} \!\!dzd\bz\, , \label{c1}
\ee
and we will see that in this case the location of the string cores may be
at orbifold singularities. From eq.(\ref{c1}) it follows that around
the eight points $z_i$ there exist a deficit angle of $\pi/2$ and it is clear
that these points cannot be thought as orbifold singularities. The deficit
angle of a fixed point of order $n$ is $2\pi(n-1)/n$. Let us assume that the
eight points in eq.(\ref{c1}) coalesced into three points of order three, three
and two, i.e.,
\bb
ds_\bot=S_2T_2U_2|\eta(S)\eta(T)\eta(U)|^4
\left|
(z-z_1)^{-3/4}(z-z_2)^{-3/4}(z-z_3)^{-1/2}\right|^{2} \!\!dzd\bz\, . \label{c2}
\ee
Around  the points $z_1,z_2,z_3$ there exist then a deficit angle of
$3\pi/2,3\pi/2$ and $\pi/2$,  respectively. This means that the transverse
space has been turned into a $T^2/Z_4$ orbifold. If  the eight points
coalesced into four points of order two each, the metric (\ref{c1})
turns out be
\bb
ds_\bot=S_2T_2U_2|\eta(S)\eta(T)\eta(U)|^4\left|\prod_{i=1}^{4}
(z-z_i)^{-1/2}\right|^{2} \!\!dzd\bz\, . \label{c3}
\ee
There exist now a deficit angle of $\pi/2$ around
each of the four points and thus the
transverse space has been turned into a $T^2/Z_2$ orbifold.
The above  configurations correspond to special points in the moduli space
with constant fields as has been discussed in \cite{dasg} for the
seven--branes of type IIB.

\section{N=2 heterotic stringy cosmic strings}

Solitonic string solutions  exist as we will see
in the N=2 heterotic theory as well.
We will explicitly construct here these heterotic N=2
four--dimensional  solutions  in the rank three, four  and five models.
These models can be obtained by reduction of N=1
supergravity coupled to N=1 super Yang--Mills in six dimensions on a
torus.
There are no Wilson line moduli for the rank three and four models
while for the rank five model there exists a
single Wilson line moduli.

\subsection{The rank three $S-T$ model}

 We will consider first the  rank three model. In this case, the vector
multiplets contain the dilaton S and the $T$
modulus which parametrize the coset $O(2,1)/O(2)$. The classical  T-duality
group  is $SL(2,\bZ)$. At a generic point of the $T$--moduli
space the gauge group is $U(1)^3$ while at $T=i$, two extra vector multiplets
become massless leading to an enhanced gauge group $U(1)^2\times SU(2)$.
One should expect then that the solitonic string solutions will  be given
by an expression similar to
eq.(\ref{solSTU}) with $U$ constant.
However, one should take into account that the classical moduli space
for the N=2 case receives quantum corrections.
Moreover, the S--duality group
$SL(2,\bZ)_S$ is not expected to be a symmetry of the full quantum theory.

The classical prepotential and the K\'ahler
potential for the model we
consider are
\bb
{\CF}^{(0)}&=&\frac{1}{2}ST \, , \label{pre0}\\
K^{(0)}&=&-\log(S-\bS)-2\log(T-\bT)^2\, . \label{kah0}
\ee
If  the above expressions were exact
one might proceed in the
construction of the solitonic strings as in sect. $2.3$.
 In the N=2 case, however, the
classical prepotential and consequently,
the K\"ahler potential, receives quantum corrections, both perturbatively and
non-perturbatively. Here we will consider only perturbative corrections
and the solution we will construct will be perturbatively exact. Due to
the N=2 non-renormalization theorems, there exist only one--loop corrections to
the classical prepotential, denoted by $h(T,U)$, as well as to the K\"ahler
potential. In fact, the classical expressions
(\ref{pre0},\ref{kah0}) are modified by quantum corrections and they  turn out
to be
\bb
{\CF}&=&\frac{1}{2}ST+h(T)+\cdots \, , \label{pre}\\
K&=&-\log(S-\bS-V_{GS})-2\log(T-\bT)  \, . \label{kah}
\ee
where the Green--Schwarz term $V_{GS}$ is
\bb
V_{GB}=
4\frac{h-\bar{h}}{(T-\bT)^2}-2\frac{\p_T h+\p_\bT \bar{h}}
{T-\bT}\, ,
\ee
and  the dots in eq.(\ref{pre})
refer to exponentially suppressed non-perturbative corrections.
In addition, the requirement  that
the $SL(2,\bZ)$ T--duality transformation $T\rightarrow
\frac{aT+b}{cT+d}$ is a K\"ahler transformation, implies that
both the dilaton $S$ and $h(T)$ transforms as
\bb
h(T)&\rightarrow& \frac{h(T)}{(cT+d)^4}+\frac{B(T)}{(cT+d)^4}\, , \nonumber \\
S&\rightarrow&S-\frac{1}{3}\p_T^2B+
2c\frac{\p_Th+\p_TB}{cT+d}-4c^2\frac{h+B}{(cT+d)^2} + const.
\ee
where $B(T)$ is at most a quartic polynomial in $T$ with real coefficients
\cite{kl},\cite{and}.
Then,  the K\"ahler potential in eq.(\ref{kah}) transforms  under $SL(2,\bZ)$
as
\bb
K\rightarrow K+2\log(cT+d)+2\log(c\bT+d) \, , \label{kahltra}
\ee
which is indeed a K\"ahler transformation.

As long as S--duality is not expected to be a symmetry of the $N=2$ string,
solitonic string solutions can be constructed by fixing the dilaton S--field to
some constant value and employing the $SL(2,\bZ)$ T--duality group for the T
modulus. Then, following the general discussion of sect. $2.1$,
we find that the metric of
the N=2 solitonic string of the rank three model is
\bb
ds^2=-dt^2+dx_3^2+(S_2-2iV_{GS})T_2^2|\eta(T)|^8\left|\prod_{i=1}^{n}
(z-z_i)^{-1/6}\right|^2dzd\bz \, , \label{met3}
\ee
where $\eta(T)$ is the Dedekind's
$\eta$--function. Near the string core we expect that
\bb
T\sim \frac{1}{2\pi i}\log (z-z_i), \ee
and thus,  $q\sim(z-z_i)$.
One may then verify that the metric is modular invariant and
has no zeroes in the complex plane. The energy on the other hand is finite and
for a single string configuration it is twice the energy of the
corresponding N=4 stringy cosmic string, i.e.,
$$ E=n\frac{\pi}{3}. $$
As a result, twelve  strings compactify the transverse space on ${\bf CP}^1$
providing a consistent  vacuum configuration.

\subsection{The rank four $S-T-U$ model}

Let us now consider the rank four N=2 heterotic $S-T-U$ model.
The moduli space in this case is
$$\frac{SL(2,\bR)}{U(1)}\times \frac{O(2,2)}{O(2)\times O(2)}, $$
where the first factor is the moduli space of the S--dilaton
and the second factor for the $T$ and $U$ moduli of the internal  torus. The
classical T--duality group is in this
case $O(2,2,\bZ)=SL(2,\bZ)_T\times SL(2,\bZ)_U$
modulo $T\leftrightarrow U$ interchange.
The classical prepotential and the K\"ahler potential for the
$S-T-U$ model  are
\bb
{\CF}^{(0)}&=&-STU \, , \label{cpre} \\
K^{(0)}&=&-\log(S-\bS)-\log\left((T-\bT)(U-\bU)\right)\, , \label{K0}
\ee
where $S,T,U$ have been defined in eq.(\ref{def}).
In the N=4 case considered in sect. $2.3$, ${\CF}^{(0)}$ and
$K^{(0)}$ were  exact and together with S--
and T--duality  employed in the construction of the N=4 stringy cosmic
string.  Here however, as in the rank three model,
quantum corrections modify both the prepotential and the
K\"ahler potential which turn out to be
\bb
{\CF}&=&{\CF}^{(0)}+h(T,U)+\cdots \, , \label{corr} \\
K&=& -\log(S-\bS+V_{GS})-\log\left((T-\bT)(U-\bU)\right)\, , \label{corre}
\ee
where the Green--Schwarz terms is
\bb
V_{GS}=-2\frac{h-\bar{h}}{(T-\bT)(U-\bU)}+\frac{\p_T h+\p_\bT \bar{h}}
{(U-\bU)}-\frac{\p_U h+\p_\bU \bar{h}}
{(T-\bT)} \, .
\ee

Proceeding as before we find that
the N=2 stringy cosmic string for the $S-T-U$ model has metric
\bb
ds^2=-dt^2+dx_3^2+(S_2-2iV_{GS})T_2U_2|\eta(T)\eta(U)|^4\left|\prod_{i=1}^{n_T}
\prod_{j=1}^{n_U} (z-z_i)(z-z_j)\right|^{-1/6}dzd\bz \, , \label{met33}
\ee
where $n_T,n_U$ are the number of times the z--plane covers the fundamental
domains of $SL(2,\bZ)_T$ and $SL(2,\bZ)_U$, respectively. The $T\leftrightarrow
U$ exchange  symmetry is broken by quantum corrections and the numbers
$n_T,\, n_U$ cannot be related any more as in 2.3. The energy turns out to be
\bb
E=\frac{\pi}{6}(n_T+n_U) \, , \ee
and  the regularity of the solution requires $n_T+n_U=24$.

\subsection{The rank five  $S-T-U-V$ model}

An interesting case of string-string dualities is provided by the heterotic
string in D=10 compactified on $K3\times T^2$ which is related to a type II
string compactified on a appropriate Calabi-Yau three-fold. There exist
successful  tests of this duality for models with small number of vector
multiplets \cite{dual},\cite{lust}.
In particular, with $N_V=4$ massless Abelian vector multiplets one
is dealing with the S--dilaton and the complex fields T and U
(besides the graviphoton)
where  T and U are the torus moduli. Solitonic string solutions of this model
has been discussed in the previous chapter. Here, we will consider the case
where additional massless Wilson line moduli exist.
In the presence of p non-vanishing Wilson lines, the classical
vector multiplet moduli space  of N=2 string compactification
turns out to be  locally the special K\"ahler manifold \cite{fer},
\cite{harvey}
$$\frac{SL(2,\bR)}{U(1)} \times
\frac{O(2,2+p)}{O(2)\!\times\!O(2+p)},$$
where again the first factor is the S-field muduli space.
The classical T-duality group is $O(2,2+p;\bZ)$.
For the special p=1 case we will discuss   here,
there exist a single
Wilson line V and the moduli space is
$$\frac{SL(2,\bR)}{U(1)} \times
\frac{O(2,3)}{O(2)\!\times\!O(2+p)}.$$ The T-duality group is $O(2,3;{\bf Z})$
 which is isomorphic to $Sp(4,{\bf Z})$. A short account of its properties
are presented  in the appendix.

The loop-corrected prepotential and K\"ahler potential for the $S-T-U-V$ model
are
\bb
{\CF}&=&-S(TU-V^2)+h(T,U,V)\, ,\label{stpre}\\
K&=&-\log(S-\bS+V_{GS})-\log\left((T-\bT)(U-\bU)-(V-\bar{V})^2\right)
\, , \label{STkah}
\ee
where $h(T,U,V)$ is the one--loop prepotential and the Green--Schwarz
term $V_{GS}$ is expressed in terms of $h$ as
\bb
V_{GS}&=&\frac{(T-\bT)(h_T+\bar{h}_\bT)+(U-\bU)(h_U+
\bar{h}_\bU)+(V-\bar{V})(h_V+\bar{h}_{\bar{V}})}
{\left((T-\bT)(U-\bU)-(V-\bar{V})^2\right)}-\nonumber \\
&&\frac{2(h-\bar{h})}
{\left((T-\bT)(U-\bU)-(V-\bar{V})^2\right)^2}\, ,
\ee
where e.g. $h_T=\partial_Th$. The T--duality transformation
$\tau\rightarrow (a\tau+b)(c\tau+d)^{-1}$
is a K\"ahler transformation and the
K\"ahler potential transforms as
\bb
K\rightarrow K+\log\left(det(c\tau+d)\right)+\log\left(det(c\bar{\tau}
+d)\right)
\, , \label{tra}
\ee
where $\tau$ is defined in eq.(A.2).

Finite energy  solitonic string solutions for the $S-T-U-V$
model may be constructed by  fixing
the dilaton to some constant value and
employing the  $Sp(4,\bZ)$ T--duality group. We will allow
again the $\tau$--field to have discontinues
jumps as we go around the string. Then, near the core of the string,
we will have
\bb
T\sim\frac{1}{2\pi i}\log (z-z_i), ~~~~~~~U\sim\frac{1}{2\pi i}\log (z-z_j)
 ~~~~~~~V\sim\frac{1}{2\pi i}\log (z-z_k)\, . \label{sim}
\ee
According to eq.(\ref{metricsol}), the metric for the $S-T-U-V$ model
takes the form
\bb
ds^2=-dt^2-dx_3^2+(S_2-2iV_{GS})(T_2U_2-V_2^2)|f(T,U,V)|^2dzd\bz \, .
\label{mmm}
\ee
The function $f(T,U,V)$ will be determined by demanding
modular invariance and  no degenerate metric.
It follows from eq.(\ref{mmm}) that in order to achieve modular invariance
$f(T,U,V)$ must contain a factor which transforms as  a
modular form of weight +1  and has no zeroes in the
fundamental domain ${\cal F}_2$. The unique form with this
properties is the twelfth root of the cusp form
$\Psi_{12}$. However, although the latter
has no zeroes in ${\cal F}_2$ it might
have zeroes in the z--plane at the locations of the string core.
There we have
$$q=\exp(2\pi iT)\sim (z-z_i),~~~~~
s=\exp(2\pi iU)\sim (z-z_j),~~~~~ r=\exp(2\pi iV)\sim (z-z_k)$$
as it follows from eq.(\ref{sim}).
Although a product expression for $\Psi_{12}$
is lacking \cite{nik},  we now that \cite{igusa}
\bb
\Psi_{12}=qs+\cdots \,  ,
\ee
which may also be seen from the degeneration limit $V\rightarrow 0$ at which
\bb
 \Psi_{12}\rightarrow \Delta(q)\Delta(s) \, .
\ee
Thus, the conditions of modular invariance and non--degeneracy of the
metric give
\bb
ds^2=-dt^2-dx_3^2+(S_2-2iV_{GS})(T_2U_2-V_2^2)\left|\Psi_{12}^{}\right|^{1/6}
\left|\prod_{i=1}^{n_T}\prod_{j=1}^{n_U}|(z-z_i)(z-z_j)\right|^{-1/6}\!\!
\!\!\!\!\!dz
d\bz.
\ee
In the degeneration limit one recovers the solution of the $S-T-U$ model
eq.(\ref{met33}).
The solution for the $T,U,V$ moduli will be given as the pull--backs of the
modular invariant functions $x_1,x_2$ and $x_3$ defined in the Appendix. The
energy finally is indeed finite
\bb
E=\frac{\pi}{6}(n_T+n_U),
\ee
and  regularity demands $n_T+n_U=24$.

\section{Conclusions}

We have constructed here solitonic string solutions of four--dimensional N=4
and N=2 heterotic theories. In the N=4 case, we have considered the dilaton
with two additional moluli. By employing the S-- and T--duality
groups as well as string/string/string triallity we were able to explicitly
find string--like configurations with finite energy per unit length.
Regular solutions are provided then by twenty--four strings since in
this case the transerse space is compactified on ${\bf CP}^1$. This is closely
related with the fact that there exist elliptically fibered manifolds
with base space ${\bf CP}^1$ as for example  K3 surfaces or Calabi--Yau
three-folds which admit two elliptic fibrations. Then, the singularities
may be resolved in a Ricci--flat way,
consistently with supersymmetry \cite{g}.

The solitonic strings of the N=4 case may also be seen
from  a type II point of view. One may compactify type II theory on
a $(T^2)^3$ which gives N=8 supersymmetry in four dimensions.
By fixing the K\"ahler
structure moduli of the tori to some constant value, the only moduli which will
appear in the four--dimensional effective theory will be three complex scalars
corresponding to the complex sructures of the internal tori. In this case, one
may express the solutions in terms of elliptic curves as has been done in the
seven--brane solution in type IIB theory in ten dimensions \cite{vaf}.

We have also constructed  solitonic string solutions of N=2 heterotic
models of rank three, four and five. In the first two cases there are no Wilson
line moduli and the solutions were found by employing the T--duality groups
$SL(2,\bZ)$ and $SL(2,\bZ)\times SL(2,\bZ)$. We kept the dilaton constant as
explained already since otherwise the construction of finite energy solutions
would not be possible. We have also consider the case were a single Wilson line
moduli is present. In this case the T--duality group is $O(3,2;\bZ)$
which is isomorphic to $Sp(4,\bZ)$ and the
solitonic string solution was expressed in terms of modular forms of the
latter. In should be not that in the general case were p Wilson line moduli are
turned on, the T--duality group is $O(2+p,2;\bZ)$ and we need to know its
modular forms with no zeroes in
$O(2+p,2;\bZ)\backslash O(2+p,2;\bR)/O(2+p)\times O(2)$. These forms
are not explicitly known and a way to construct them may probably be based on
\cite{boss} where automorphic forms of $O(2+p,2;\bR)$ with well
controlled pole structure have been introduced.

Finally, it should be noted that a  single solitonic string  is  not a
consistent solution. One has to consider  multi--string configurations as
consistent string backgrounds. In this case, the transverse space is
compact and  at a generic point of the moduli space, it is a  sphere. At
special points, however, it turns into an orbifold. The location of the string
cores are then at the fixed points of that orbifold.

\vspace{2cm}

{\large\bf Appendix}\setcounter{equation}{0}
\def\theequation{A.\arabic{equation}}

Here we review some properties of  $Sp(4,\bZ)$~\cite{igusa}--\cite{dev}.
The group   $Sp(4,{\bf Z})$ is subgroup of $Sp(4,\bR)$ and consists of all
integral $4\times4$ matrices $$M=\left( \begin{array}{cc}
a&b\\c&d\end{array}\right),$$ such that $M^tJM=J$
 where $a,b,c,d$ are integral $2\times2$ matrices and
$$J=\left( \begin{array}{cc}
0&{\bf 1}_{2\times2}\\-{\bf 1}_{2\times2}&0\end{array}\right).$$
The standard action of $Sp(4,{\bf Z})$ on the Siegel upper half space
${\cal H}_2=O(2,3)/O(2)\!\times\!O(3)$ is given by
\bb
\tau\rightarrow (a\tau+b)(c\tau+d)^{-1}\, ,\label{t}
\ee
where
\bb
\tau=\left(\begin{array}{cc}
T&V\\
V&U\end{array}\right)\in {\cal H}_2,\,
\ee
with  $Im\tau=(T_2U_2-V_2^2)>0$
\footnote[1]{$T_1,T_2,U_1,U_2,V_1,V_2$
are the real and imaginary  parts of the complex fields T,U and V
respectively.} Similarly to the $SL(2,\bZ)$ group,
$Sp(4,{\bf Z})$ is generated by
$$\left(\begin{array}{cc}
0&{\bf 1}_{2\times2}\\
-{\bf 1}_{2\times2}&0\end{array}\right)
{}~~~~~~~\left(\begin{array}{cc}
A&0\\
0&A^\ast\end{array}\right),~~~~~~~
\left(\begin{array}{cc}
{\bf 1}_{2\times2}&B\\
0&{\bf 1}_{2\times2}\end{array}\right), $$
where $A\in GL(2,\bZ)$ with $A^\ast=(A^t)^{-1}$ and
$$
B=\left(\begin{array}{cc}
1&0\\
0&0\end{array}\right), ~~~~~~~
\left(\begin{array}{cc}
0&0\\
0&1\end{array}\right), ~~~~~~~~
\left(\begin{array}{cc}
0&1\\
1&0\end{array}\right)\, .
$$
The
$Sp(4,{\bf Z})$ fundamental domain ${\cal F}_2$
can be defined by the conditions
\vspace{-2mm}

\begin{enumerate}
\item $|T_1|\leq\frac{1}{2},\,  |U_1|\leq\frac{1}{2},\, |V_1|\leq\frac{1}{2}
\, , $\vspace{-2mm}

\item $0\leq|2V_2|\leq T_2\leq U_2\, , $ \vspace{-2mm}

\item $|det(c\tau+d)|\geq 1$ for all
$\left(\begin{array}{cc}a&b\\c&d\end{array}\right)\in Sp(4,\bZ)$.
\vspace{-2mm}
\end{enumerate}

A Siegel modular form F of weight k is a holomorphic function on ${\cal H}_2$
with the property
\bb
F\left((a\tau+b)(c\tau+d)^{-1}\right)=det(c\tau+d)^kF(\tau) \, .\label{k}
\ee
Such forms appear  in string theory
in two--loop amplitudes \cite{poly}
as well as in some recent developments  \cite{dev}.
Examples of such modular forms are provided by the Eisenstein series
\bb
E_k(\tau)=\sum_{c,d}\frac{1}{det(c\tau+d)} \, ,
\ee
where the summation is over all inequivalent bottom rows of elements of
$Sp(4,{\bf Z})$. One may easily prove that $E_k(\tau)$ are modular with weight
k for $k>3$.
The Eisenstein series $E_4,E_6,E_{10}$ and $E_{12}$ are algebraically
independent over $\bC$ and they generate the graded ring of
even modular forms.
Similarly to the $SL(2,\bZ)$ case, there are also cusp forms for the group
$Sp(4,{\bf Z})$ which are the forms  $\Psi_{10}$ of weight 10,
$\Psi_{12}$ of weight 12 and $\Psi_{35}$ of weight 35. In addition, one
defines $\Psi_5=\Psi_{10}^{1/2}$ and $\Psi_{30}=\Psi_{35}/\Psi_5$.

There exist also $SL(4,\bZ)$ modular functions, the counterparts of
the j--modular function. There exist three such functions which can be
written as
\bb
x_1=E_4\Psi_{10}^2/\Psi_{12}^2,
{}~~~~~x_2=E_6\Psi_{10}^3/\Psi_{12}^3,
{}~~~~~x_3=\Psi_{10}^6/\Psi_{12}^5\, .
\ee

In general the cusp
forms have zeroes on rational quadratic divisors  ${\it H}_{\ell}$
of ${\cal H}_2$ with  discriminant
$D(\ell)=\beta^2-4\delta\epsilon-4\alpha\gamma$.
${\it H}_{\ell}$ is defined as the set
$$
{\it H}_\ell=\{\left(\begin{array}{cc}T&V\\V&U\end{array}\right)\in {\cal H}_2
| \alpha+ \beta T+\gamma V+\delta U +\epsilon (V^2-TU)=0\}\, ,
$$
where $\ell=(\alpha,\beta,\gamma,\delta,\epsilon)\in \bZ^5$ is primitive, i.e
their  greatest common divisor is one. The divisor ${\it H}_{\ell}$  exists
if $D(\ell)>0$ and it  determines
the Humbert surface ${\it H}_{D}$ in $Sp(4,{\bf Z})\backslash{\cal H}_2$ which
is the union of all divisors of discriminant $D(\ell)$. The divisors of
$\Psi_{10}$ and $\Psi_5$ are the Humbert surface
$${\it H}_1=\{\tau\in Sp(4,{\bf Z})|\tau=
\left(\begin{array}{cc}T&0\\0&U\end{array}\right)\},$$
the divisors of $\Psi_{30}$ are the surface
$${\it H}_4=\{\tau\in Sp(4,{\bf Z})|\tau=
\left(\begin{array}{cc}T&V\\V&T\end{array}\right)\},$$
and the divisors of $\Psi_{35}$ is the union of ${\it H}_1$ and ${\it H}_4$.
On the other hand, the unique cusp form without divisors is $\Psi_{12}$.

Finally, let us mention that in the degeneration limit $V\rightarrow0$
\bb
E_4&\rightarrow& G_4(T)G_4(U) \, ,\nonumber \\
E_6&\rightarrow& G_6(T)G_6(U) \, , \nonumber \\
\Psi_5&\rightarrow &0\, , \nonumber \\
\Psi_{12}&\rightarrow& \Delta(T)\Delta(U) \, , \nonumber \\
\Psi_{35}&\rightarrow&  \Delta(T)^{5/2}\Delta(U)^{5/2}
\left(j(T)-j(U)\right)\, , \label{limit}
\ee
where $\Delta=\eta^{24}$ is the $SL(2,\bZ)$ cusp form and $G_4,G_6$ are the
Eisenstein series of $SL(2,\bZ)$ of weight four and six, respectively.

\end{document}